\documentclass[twocolumn,aps,prl,amsmath,showpacs]{revtex4}
\usepackage{graphicx,dcolumn,bm,amssymb,amsmath}
\usepackage{color}

\def\be{\begin{equation}}
\def\ee{\end{equation}}

\begin{document}

\title{How cells feel: stochastic model for a molecular mechanosensor}
\author{Matteo Escud\'e, Michelle K. Rigozzi and Eugene M. Terentjev}
\affiliation{Cavendish Laboratory, University of Cambridge, JJ Thomson Avenue,
Cambridge CB3 0HE, U.K.}
\date{\today}

\begin{abstract}
\noindent  Understanding mechanosensitivity, i.e. how cells sense the stiffness of their environment is very important, yet there is a fundamental difficulty in understanding its mechanism: to measure an elastic modulus one requires two points of application of force - a measuring and a reference point. The cell in contact with substrate has only one (adhesion) point to work with, and thus a new method of measurement needs to be invented. The aim of this theoretical work is to develop a self-consistent physical model for mechanosensitivity, a process by which a cell detects the mechanical stiffness of its environment (e.g. a substrate it is attached to via adhesion points) and generates an appropriate chemical signaling to remodel itself in response to this environment. The model uses the molecular mechanosensing complex of latent TGF-$\beta$ attached to the adhesion point as the biomarker. We show that the underlying Brownian motion in the substrate is the reference element in the measuring process. The model produces the closed expression for the rate of release of active TGF-$\beta$, which depends on the substrate stiffness and the pulling force coming from the cell in a subtle and non-trivial way. It is consistent with basic experimental data showing an increase in signal for stiffer substrates and higher pulling forces. In addition, we find that for each cell there is a range of stiffness where a homeostatic configuration of the cell can be achieved, outside of which the cell either relaxes its cytoskeletal forces and detaches from the very weak substrate, or generates an increasingly strong pulling force through stress fibers with a positive feedback loop on very stiff substrates. In this way, the theory offers the underlying mechanism for the myofibroblast conversion in wound healing and smooth muscle cell dysfunction in cardiac disease.
\end{abstract}

\maketitle

Complex living organisms are made of trillions of cells that constantly grow, interact and reproduce, each one of them performing specific and fundamental tasks for life. Scientists have always been fascinated by how each one of them appears to be aware of its role and able to interact with its surroundings as if they possessed true sensory organs. The general question we seek to answer in this work is: how do cells feel the mechanical stiffness of their environment? Through the application of a classical physical model of coupled stochastic processes we expand our understanding of the sensory activity of cells.

A large amount of research has been devoted to studying the mechanisms behind chemical signal detection at the cellular level, a sensory activity that could be compared to the olfactory ability of animals and humans. However, in the last two decades there has been growing evidence that cells are not only receptive to the concentration of chemicals in their environment, but also to mechanical properties of objects they are in contact with. These mechanical properties include viscosity, hydrostatic pressure and deformability, as well as topographic profile: features of materials that animals perceive through complex tactile organs. To provide concrete examples: cellular cytoskeletal changes~(1-3), adhesions~(4), contractile forces~(5), stiffness~(2) and migration~(4) have all been shown to be affected by the mechanical properties of the substrate the cell is attached to. Cellular differentiation, based on changes in substrate stiffness, is another emblematic area where mechanosensitivity is observed. Notable evidence for this includes the work by Engler et al.~(6) on how stem cell lineage is specified by matrix elasticity. A similar example, which is a more specific focus of this paper, is the process of differentiation of fibroblasts into myofibroblasts as a response to a stiffer substrate. The most remarkable biological investigation of this system has been carried out by Wipff and coauthors~(7).

Mechanosensing is defined as when at least one chemical reaction in the cell changes characteristics in response to a change in its mechanical environment. We specifically wish to distinguish from a very different case when external forces are applied to the cell. There is a lot of research on that problem, but ultimately it has no mystery: there are many molecular processes that respond to an applied force and probably several different ones are employed by cells in different situations. It is much more challenging to understand how cells could detect the stiffness of a passive matrix they are in contact with. Such a sensor requires a mechanical structure with the ability to produce changes in a chemical signal, that are in turn detectable by receptors in the cell. The sensor specifically investigated in this work is the latent complex of the transforming growth factor-$\beta$ (latent TGF-$\beta$): a large biomolecular complex which has been shown to be associated with smooth-muscle actin expression~(8) and, in particular, to have the stretch-sensing features that cause the embryonic stem cell to smooth muscle cell~(9) or the fibroblast to myofibroblast differentiation~(7).

We defer the biological details to the next section. What is specifically relevant to the basic physical picture is the fact that a passive biomolecular complex is able to perform a measurement of stiffness on its environment. When humans wish to detect the stiffness of an object, with machinery or with their tactile organs, they usually perform two separate measurements: force or stress; and displacement or strain. The measure of ``stiffness'' is obtained by a ratio relating these two physical parameters. What is crucial to this kind of measurement is the fact that the apparatus requires two points of application: to measure displacement a fixed reference point is always needed; to measure force one has to ensure a reference point as well. We need \underline{two} fingers to squeeze a test object from two sides -- or use just one finger for measurement of an object against a reliably fixed wall. In the same way, all engineered devices ultimately have two points of application on the test sample. A simple sensor like the latent TGF-$\beta$ complex, which is effectively a molecular protuberance outside the cell surface, does not possess these features: it only has one `finger' (or point of application) through which a pulling force is transmitted from the cell interior and no {\it a priori} information about the reference point in the substrate, so how can it perform this measurement? This is the question driving our work.

As a result of this work we conclude that the measuring process is carried out thanks to the microscopic thermal fluctuations (Brownian motion) of the sensor and of the substrate, which result in a  different statistical behavior in environments with different stiffness. The biomolecular measuring complex has evolved so that its characteristic energy scale is comparable to that of thermal fluctuation in the typical cell environment, allowing the small cellular forces to affect the statistics of latent complex rupture.

Other physical models have been developed in this field, notably Schwarz et al.~(10) have proposed a model that is meant to describe the mechanosensing activity of focal adhesions. Specifically, they introduce a two-spring model where the extracellular and intracellular mechanical environments are described as two springs in series connected by a breakable bond: a construction very similar to what we shall be employing below. However, in keeping with the established methods, they assume that the cell imposes a force-dependent velocity $v(F)$ on the first spring, while the far end of the second spring remains fixed. This makes the whole problem dynamic and the fundamental difficulty of a mechanical measurement with just one point of application is no longer present: there is never a state of mechanical equilibrium along the series of springs. The force-dependent rate of rupture is then written by using the phenomenological expression of Bell: $k(F) = k_0e^{Fx_0/k_BT}$, where $k_0$ is the rate of rupture without force applied and $x_0$ is the characteristic length scale describing the free energy of the bond~(11). Of course, the Bell formula has been originally derived by Kramers~(12), in extending his classical solution for the equilibrium rate of escape over an energy barrier to when an external force modifies the potential energy:  $V(x)-Fx$. These ideas were extensively used by Dudko and coauthors~(13) in the context of single-molecule pulling experiments.

There are a number of elegant experiments that demonstrate the constant nature of the force exerted by the cell~(14-16). Compared to Schwarz et al.~(10), we offer a more fundamental explanation of the physical phenomenon as it draws on the principles of statistical mechanics rather than on a phenomenological observation, and also observes the mechanical balance laws.  Our work also provides an improvement to the Dudko et al. model~(13) as, approaching the critical breakdown force, our expression for the rate of escape shows the correct diverging behavior. The final result we obtain has the Bell formula in the high barrier limit embedded in it, however, other factors critically involving the stiffness of the substrate dominate the behavior.

\begin{figure} 
\centering
\includegraphics[width=0.45\textwidth]{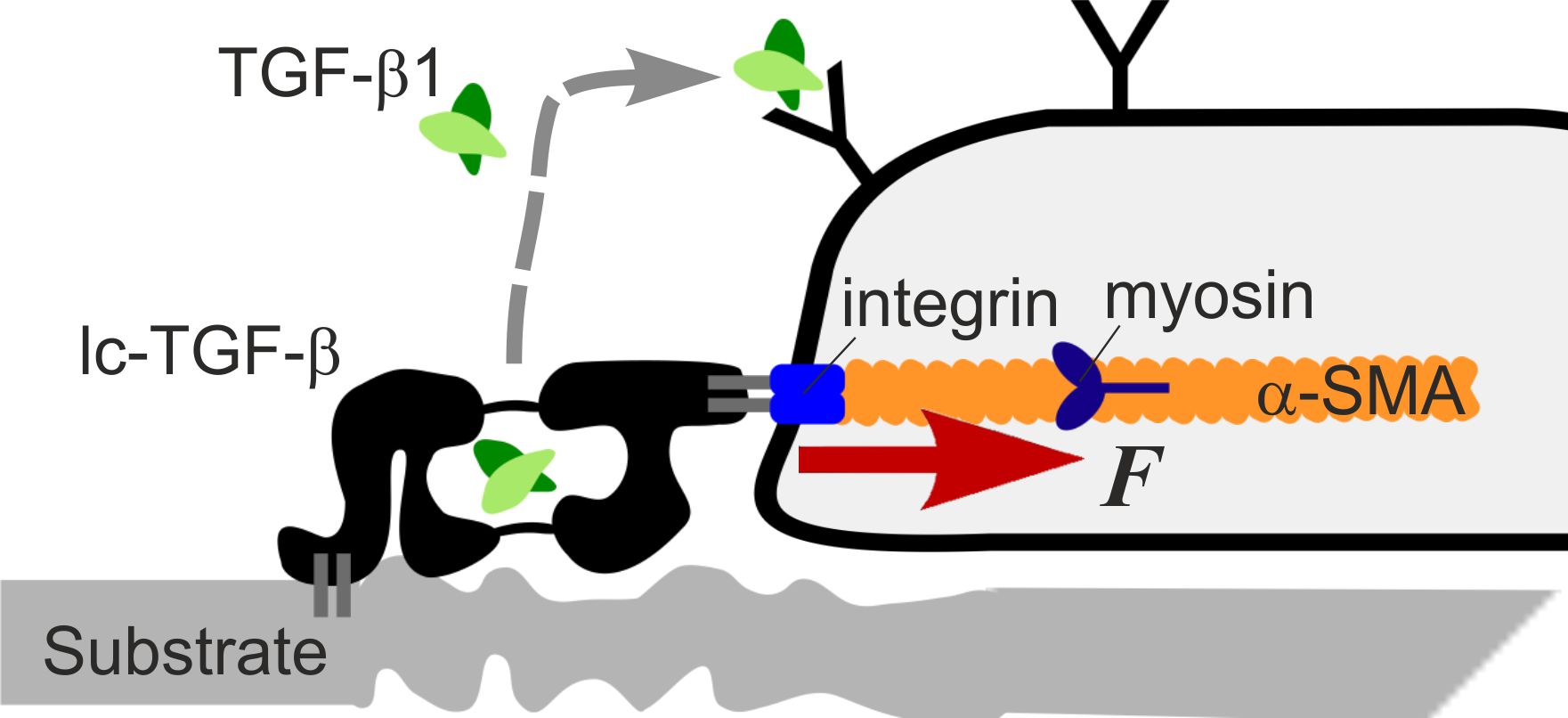}
\caption{Illustration of the biological system: the cell attachment to a deformable substrate, the application of pulling force $F$ and the release of active TGF-$\beta$. On binding to a receptor on the cell surface, TGF-$\beta$ initiates two processes -- of the production of new latent complexes (to replenish the broken ones) and of smooth-muscle actin (to stiffen the stress fibers and increase the pulling). Although this image is frequently found in the literature, it is not often appreciated that the force acting on the latent complex is always $F$ irrespective of how much the substrate is deformed.} \label{fig:pulling}
\end{figure}
\section{The biological system}\label{The Biological System}
The specific stretch-sensor base dthe latent complex of TGF-$\beta$ has been selected because it is expected to be representative of a broader family of mechanosensors and also because of the abundance of biophysical research providing reliable evidence of its behavior. In this account of the biological system we primarily follow~(1,7,17).

Fibroblasts are the most common cell in connective tissue and among other functions are crucial to the process of physiological tissue repair. Specifically, when a wound is formed, fibroblasts differentiate into myofibroblast, a highly contractile form that helps to pull the wound together and facilitates the healing process. The highly contractile state of the myofibroblasts is generated by $\alpha$-smooth muscle actin ($\alpha$-SMA) in stress fibres, appearance of which marks the transition between the two states.  In a very similar way, smooth muscle cells of arterial wall stiffen by over-expression of $\alpha$-SMA when a rigid crystalline plaque forms underneath them in the arterial cavity. In both cases the release of TGF-$\beta$ has been unambiguously registered~(7,8).

The differentiation of fibroblasts into myofibroblasts has been shown to be reproducible \emph{in vitro} by plating fibroblasts on substrates with different stiffness. For substrates stiffer than a specific Young modulus, $E \gtrsim $ 11kPa~(7), fibroblasts differentiate into myofibroblasts.  If plated on a softer substrate ($E \ll$ 11kPa), however, fibroblasts do not transform and remain in the fibroblast state -- and myofibroblasts turn back into fibroblasts. The value around 10kPa is particularly interesting as it is similar to the difference in stiffness between wound tissue and typical connective tissue~(18).

Strong evidence suggests that this phenotype change is caused by the mechanical activation of latent TGF-$\beta$ as a biomarker~(19). When active TGF-$\beta$ is released into the cell surroundings, it can bind to specific receptors on the cell surface, and the cell is induced to synthesize three types of protein crucial to the signalling loop: more of the latent TGF-$\beta$ complex, $\alpha$-SMA, and matrix proteins~(7,20). The latent complex is produced to add to the body of sensors, some of which have been broken, and the $\alpha$-SMA forms contractile fibres in the cell, increasing the force applied to the latent complex molecules. Hinz et al.~(20) specifically confirm that an increased $\alpha$-SMA expression is sufficient to enhance fibroblast contractile activity.
The latent complex adheres to integrins on the cell surface which, intracellularly, are bound to the contractile cytoskeleton. Outside the cell, the latent complex is attached to the extracellular matrix via proteins such as fibronectin~(22), fig.\,1. When the cell contracts, the latent complex is put under tension. However, we have to question the beautiful and intuitively clear picture proposed by Wipff et al.~(7,21): since the cell exerts a (constant) force, it does not matter whether the substrate deforms (in the soft case) or stays rigid -- basic mechanics tells that in all cases the tension remains constant along the chain of elements in fig.\,1. If one wants to relate the rate of active TGF-$\beta$ release to the force applied to the latent complex, e.g. by Bell's formula, as in~(10) -- this cannot be different between the soft and stiff cases. Resolving this problem is our aim here.

\section{The model}\label{The Model}
No analytical theory for the mechanosensing role of TGF-$\beta$ exists in the literature. Our model is inspired by the two-spring construction of Schwarz et al.~(10) and is illustrated in fig.\,2.
We treat the substrate (which could be the extracellular matrix, ECM, or an artificial material) as a one-dimensional viscoelastic Voigt system: the effective spring attached to a distant fixed point is characterized by two parameters, an effective spring constant $\kappa$, representing its elastic modulus, and a damping coefficient $\gamma_1$. It is important to distinguish this from a commonly used Maxwell model for viscoelasticity of polymer melts and complex fluids: although the relaxation time is the same, only with an elastic and a dissipative elements in parallel can one model a gel with an equilibrium elastic modulus.

 This viscoelastic spring is
 connected in series with the latent complex of TGF-$\beta$, which in turn is connected to the rigid integrin complex in the cell that applies a constant pulling force $F$ to the system. An assumption of constant local force $F$ exerted by the cell is adopted because for the timescales relevant to thermally driven escape, the force may be assumed time-independent. Any change in pulling force by the cell is caused by the chemical signalling loop, which is dependent on diffusion distances much larger than the ones characterizing the model.

 The latent TGF-$\beta$, is modeled by a lock, which could break if it gains enough thermal energy to ``escape'' a barrier given by a potential $U(x)$, meaning the rupture of the complex and release of the small molecule responsible for the signal: the active TGF-$\beta$. The sketch in fig.\,2 shows that it is the relative displacement of the two ends ($x_2-x_1$) that is the correct argument of the potential. The damping properties of the lock are described by a coefficient $\gamma_2$, in the same way as with the viscoelastic spring for the substrate. Both the lock and the substrate are subject to distinct stochastic forces, satisfying their separate fluctuation-dissipation relations, which (the thermal motion) is the key element to the whole process. We write and solve a set of coupled stochastic differential equations that describe the system to find the rate of escape over such a barrier. This is the rate at which active TGF-$\beta$ is released, initiating the signalling loop that will ultimately cause, e.g. the over-expression of $\alpha$-actin resulting in the fibroblast conversion to myofibroblast, or the stiffening of smooth muscle cells.

\begin{figure}[t]
\centering
\includegraphics[width=0.32\textwidth]{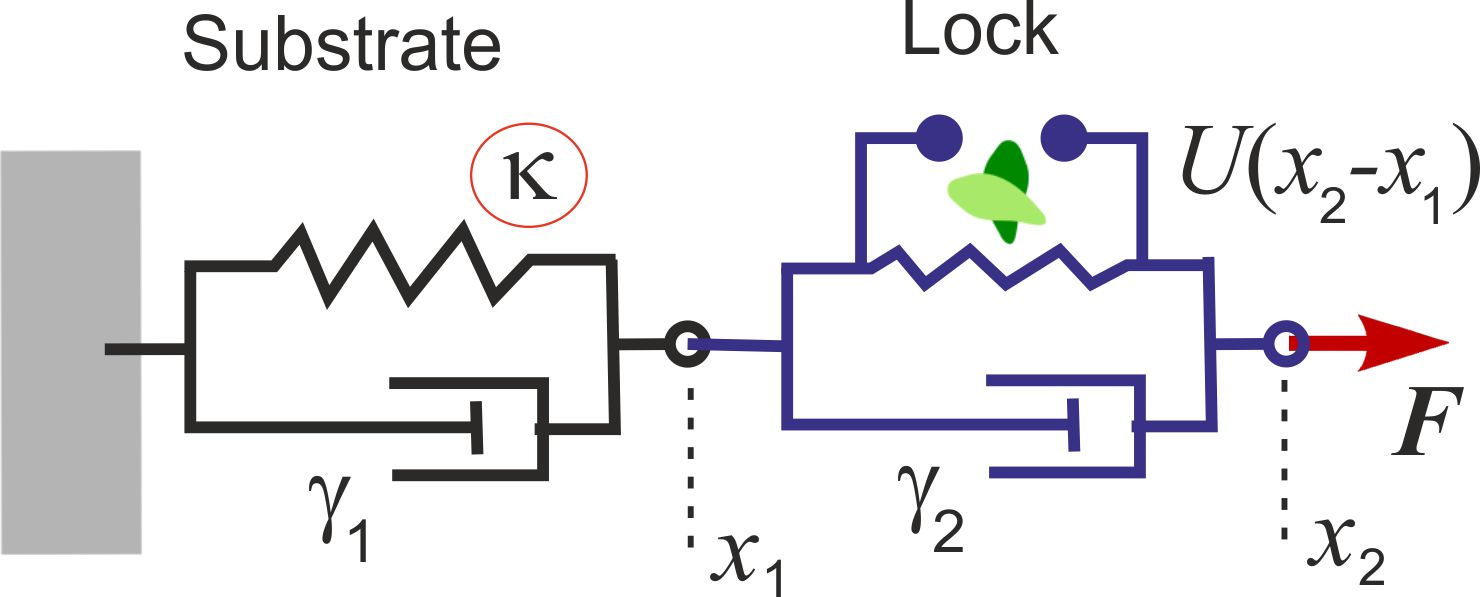}
\caption{The ``lock'', representing the latent TGF-$\beta$ complex, is pulled by the cell with a constant force $F$ and is coupled to a viscoelastic substrate with the equilibrium stiffness $\kappa$. Both elements have a dissipative (friction) component to their response, described by the constants $\gamma_1$ and $\gamma_2$, respectively. The positions of the substrate and lock deviation of equilibrium are $x_1$ and $x_2$. }
\label{fig:model}
\end{figure}

To write the master kinetic equation for the system we shall follow a route similar to the one taken by Graham~(23), as it provides a concise and rigorous method to derive the Fokker-Planck equation from a set of coupled overdamped Langevin equations:
\begin{eqnarray}
\gamma_1 \dot{x}_1 &=&  - \kappa x_1 + \frac{\mathrm{d}U}{\mathrm{d}(x_2-x_1)}  + \sqrt{2k_BT\, \gamma_1} \cdot \zeta_1(t), \\
\gamma_2 \dot{x}_2 &=& - \frac{\mathrm{d}U}{\mathrm{d}(x_2-x_1)} + F + \sqrt{2k_BT\, \gamma_2} \cdot \zeta_2(t),
\end{eqnarray}
where the base stochastic process $\zeta_i(t)$ is assumed to be Gaussian and normalized to unity, summation over repeated indices is implied. Note that it is the difference in independent position coordinates $x_2-x_1$, that affects the lock.
    The steps of derivation are standard, and as a result we obtain the Cartesian components of diffusion current in the space of $\{x_1, x_2 \}$:
\begin{equation} \label{eq:Ji}
J_i = - \frac{k_BT}{\gamma_i} e^{-V_{\rm eff}/k_BT} \frac{\partial}{\partial x_i} \left(e^{V_{\rm eff}/k_B T} P \right) \qquad i=1,2 \;,
\end{equation}
where $V_{\rm eff}(x_1,x_2)= \frac{1}{2} \kappa x_1^2 - Fx_2 + U(x_2-x_1) $
represents the effective potential landscape over which the substrate and the latex complex move, subject to thermal excitation. The components $J_1$ and $J_2$ naturally have different diffusion constants: $k_BT/\gamma_1$ and $k_BT/\gamma_2$, respectively. This issue is addressed by changing variables:
$x_1 \rightarrow \sqrt{{\gamma_2}/{\gamma_1}} \tilde{x}_1, \ x_2 \rightarrow \sqrt{{\gamma_1}/{\gamma_2}} \tilde{x}_2$, giving the unique diffusion coefficient $\tilde{D} = {k_BT}/\sqrt{\gamma_1 \gamma_2}$ in the transformed Eq.3. We now have an expression for the vector current in two dimensions and can apply the Kramers theory of escape over a potential barrier, following the original work (12) and the analysis by (24). Interested readers may obtain more technical details in the Supporting Information.

\section{The potential landscape}\label{Analysis of the Potential Surface}
To study the effective potential landscape it is most convenient to transform the coordinates, replacing the second variable $x_2$ with the net separation $u=x_2-x_1$, which is the natural variable of the lock potential $U(u)$. To be specific,  it is useful to express $U(u)$ in an explicit functional form. Following Dudko et al.~(13) we use a cubic function, which gives a realistic profile to the barrier and naturally excludes the possibility of rebinding by falling to negative infinity for large $u$. It is tuned so that the minimum is at $u=0$ and the maximum of height $\Delta$ at $u=u_0$:
\begin{equation}
U(u) = \frac{3}{2} \Delta \left( \frac{u}{u_0} - \frac{1}{2} \right) - 2\Delta \left(\frac{u}{u_0} - \frac{1}{2} \right)^3 + \frac{\Delta}{2}.
\end{equation}
Substituting this into $V_{\rm eff}(x_1,u)$, the potential landscape may be readily plotted, for instance at constant $x_1=0$ (i.e. non-deformed substrate), to study how the landscape evolves for different values of force $F$ and elastic constant $\kappa$, fig.\,3(a). Contour plots of the effective potential as a map on ($x_1,u$) plane provide a useful tool for finding the optimal path of the system evolution, fig.\,3(b).

\begin{figure} 
\centering
\includegraphics[width=0.4\textwidth]{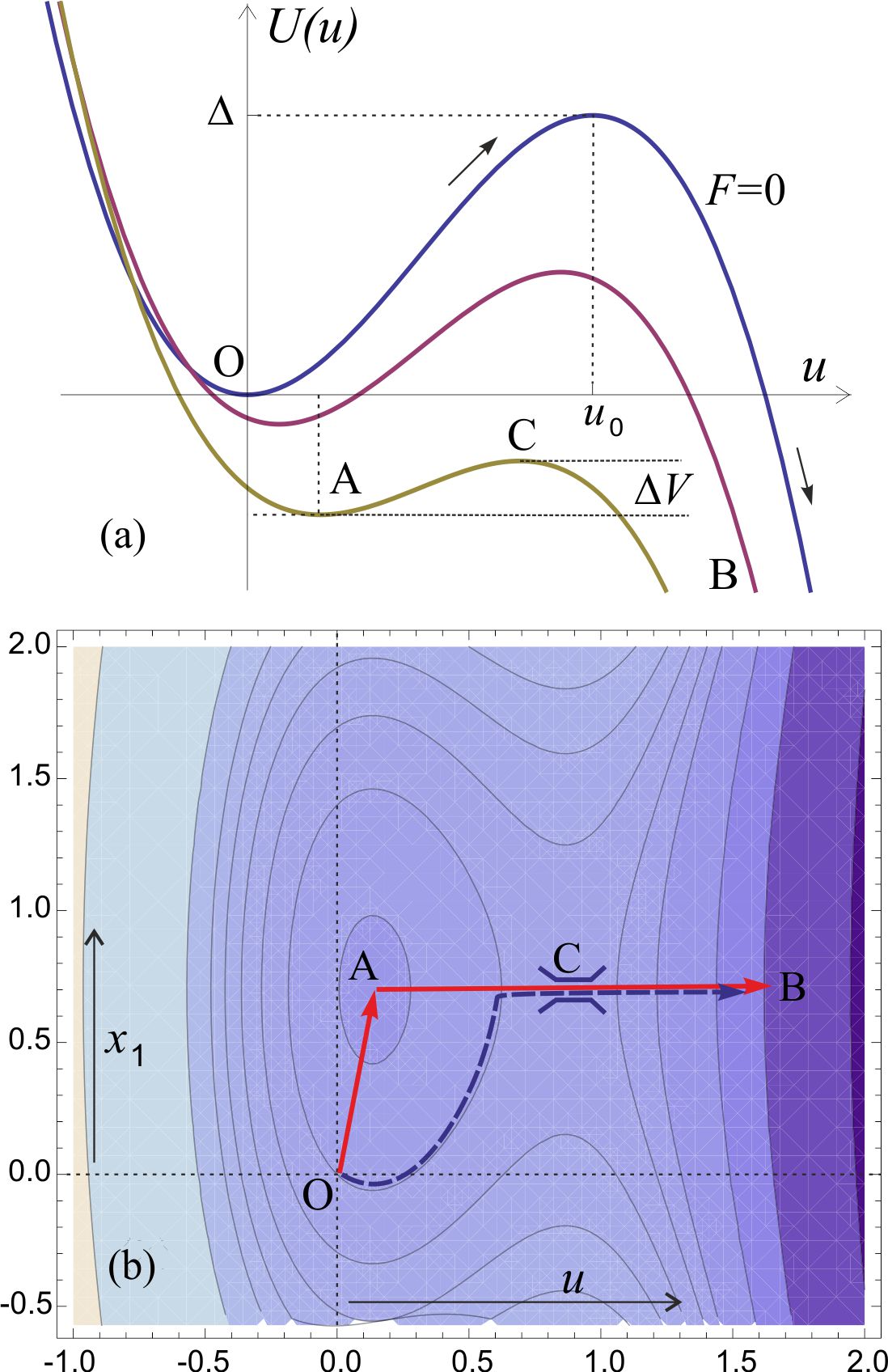}
\caption{(a) Plot of the cubic function describing $V_{\rm eff}$ at $x_1=0$, the bottom of the well is at $u=0$ and the top of the energy barrier of height $\Delta$ at $u=u_0$. The subsequent dashed lines are for the increasing applied force $F$, labeled on plot. (b) Contour plot of $V_{\rm eff}(x_1,u)$; darker areas represent lower energy values. In this plot we have chosen the values of $\widetilde{\kappa}=\kappa u_0^2/\Delta = 0.4$ and $\widetilde{f}= Fu_0/\Delta = 0.3$. The latter value corresponds to the $0.2$ of the critical force $F_C$. The solid arrows outline the approximation used to the preferred path O $\rightarrow$ A $\rightarrow$ C $\rightarrow$ B along the potential surface, while the dashed line gives an example of another path, with lower potential barrier but an effectively much lower rate.} \label{fig:cubic}
\end{figure}

We now proceed to analyze the potential surface more rigorously by finding the position of the extrema and the values of $V_{\rm eff}(x_1,u)$ at those points. By calculating the derivatives of the function one finds the potential has two extrema (points A and C in fig.\,3):
\begin{eqnarray}\label{eq:extrema}
 && x_1(A) = \frac{F}{\kappa}, \ \  u(A,C)=\frac{1}{2} u_0 \left( 1 \mp \sqrt{1- \frac{2}{3} \frac{Fu_0}{\Delta}} \right)  \\
 && V_{\rm eff}(A) \approx  -\frac{F^2}{2\kappa } - \frac{F^2u_0^2}{12\Delta} , \ \  \Delta V =\frac{\Delta (3-2Fu_0/\Delta)^{3/2}}{3\sqrt{3}}  \nonumber
\end{eqnarray}
The bottom of the well (A) and the saddle point (C) have the same $x_1$ coordinate. For $F_C=3{\Delta}/2{u_0}$ the two extrema coincide and the path from the origin to the bottom of the potential landscape has no energy barrier: the system shown in fig.\,2 breaks down. This effective barrier has to be overcome by the substrate-and-lock system as it  evolves under thermal fluctuations. The important point in this potential landscape analysis is that the barrier height $\Delta V = V_{\rm eff}(C)-V_{\rm eff}(A)$ does not depend on the substrate stiffness, while the depth of the potential well (A) does include the constant factor $-\beta F^2/2\kappa$.

\section{The rate expression}

Applying the Kramers theory to generalized problems of thermally-activated diffusion over a potential barrier is a relatively standard operation (25) and we are not going to go into details. There have been more recent papers on the calculation of escape rate in 2-dimensions (26,27), which often are variations the original treatment by Langer (28).  However, there are two important new steps that we make, which will be discussed here (see the Supporting Information for technical details).

First of all, we want to be able to have solutions for any value of external applied force $F$, including near the critical breakdown point $F_C$. The standard approach, e.g. (13,26), is to remain in the limit of weak forces ($F \ll \Delta/u_0$ in our notation) where the required integration over the potential barrier ($\int_A^B \exp [V_{\rm eff}(u)/k_BT] du$) can be accurately approximated by the saddle-point expression at point (C). However, this approximation fails as the barrier height $\Delta V$ diminishes to zero and the predicted reaction rate no longer describes the correct physical process of free escape. We use a different method, essentially approximating the whole exponential (as opposed to the exponent) by a parabola pinned to the correct values of $V_{\rm eff}$ at (A) and (C) and integrate only from A to C, doubling the value of the integral afterwards. This method accurately describes the result near the critical breakdown point (A $\rightarrow$ C), and retains the correct activation exponential factor at smaller forces. In effect, this method makes the same-magnitude error in evaluating the integral, but retains the accuracy near the critical force point.  As a result, the steady-state flux of escape over the barrier (C) takes the form:
 \begin{equation} \label{eq:Jfinal}
J \approx \frac{3}{2}\frac{\tilde{D} }{(u_C-u_A) \left(1+2 \mathrm{e}^{\Delta V /k_BT} \right) }  \cdot P(A) ,
\end{equation}
with $u_C, \, u_A$ and $\Delta V $ (all functions of $F$) are given in Eqs.5 and $P(A)$ is the probability for the system to reach the bottom of the potential well (A).

Second point of novelty is related to the evaluation of the number of particles available at (A), $N_A$, that is, before the barrier -- so that the rate of the escape over the barrier is given by the ratio $k=J/N_A$ (as is standard in the Kramers problem). This number is given by the integral around this potential minimum: $N_A= \int P(A) \exp [-V_{\rm eff} /k_BT] \mathrm{d} l$. As we are on a 2-dimensional plane -- the path of this integration is not straight on the $(x_1,u)$ plane. The path is marked on the landscape map in fig.\,3(b) and the curvature of energy surface is not the same in the $x_1$- and $u$-directions around the minimum at (A). On the first leg, when the system is `sinking' down the potential well (O$\rightarrow$A) -- the curvature is $\kappa$, the elastic modulus of the substrate, while on the second leg, when the system is approaching the saddle (A$\rightarrow$C) -- the curvature is $6(\Delta/u_0^2)\sqrt{1-2Fu_0/3\Delta}$. Therefore the overall result for the number of particles is determined by the weighted mean of the two curvatures (due to the Gaussian nature of the integral), giving
\begin{equation}\label{eq:nA}
N_A \approx  \sqrt{2\pi k_BT}  \,\mathrm{e}^{F^2/2\kappa k_BT} P(A) \sqrt{\frac{F + \kappa u_0 \sqrt{1-2Fu_0/3\Delta}}{3\kappa (2\Delta/u_0 - F) }} \ .
\end{equation}

Note that the dependence on external pulling force $F$ is prominent in the expression for the flux over the barrier, Eq.6, entering via $(u_C-u_A)$ and $\Delta V$ -- both shown in Eqs.5. In contrast, it is the statistical weight of the system at the bottom of the deformed potential well, Eq.7, where a strong dependence on the substrate elasticity $\kappa$ appears. The resulting expression for the rate of escape (the rupture of the latent TGF-$\beta$ complex, in our context) is given by the ratio of the two expressions (Eqs.6 and 7), $k=J/N_A$; it is quite a complicated and cumbersome formula. Here, for simplicity and clarity, we only plot the rate dependence on the substrate stiffness and also examine two relevant limiting cases.

Firstly, to be able to plot functions in non-dimensional form, we need to establish the relevant scales. Let us compare the applied force to the strength of the original lock potential $U(u)$, and assume that in most cases the force is weaker than the native latent complex (i.e. does not break it outright): this means small $\widetilde{f}= Fu_0/\Delta$. Let us also compare the stiffness of the substrate to the stiffness of the original lock, that is, measure $\widetilde{\kappa}=\kappa u_0^2/\Delta $. We must always assume that the original latent complex is stable at the typical temperature, that is, $\Delta / k_BT$ is large.

\begin{figure} 
\centering
\includegraphics[width=0.45\textwidth]{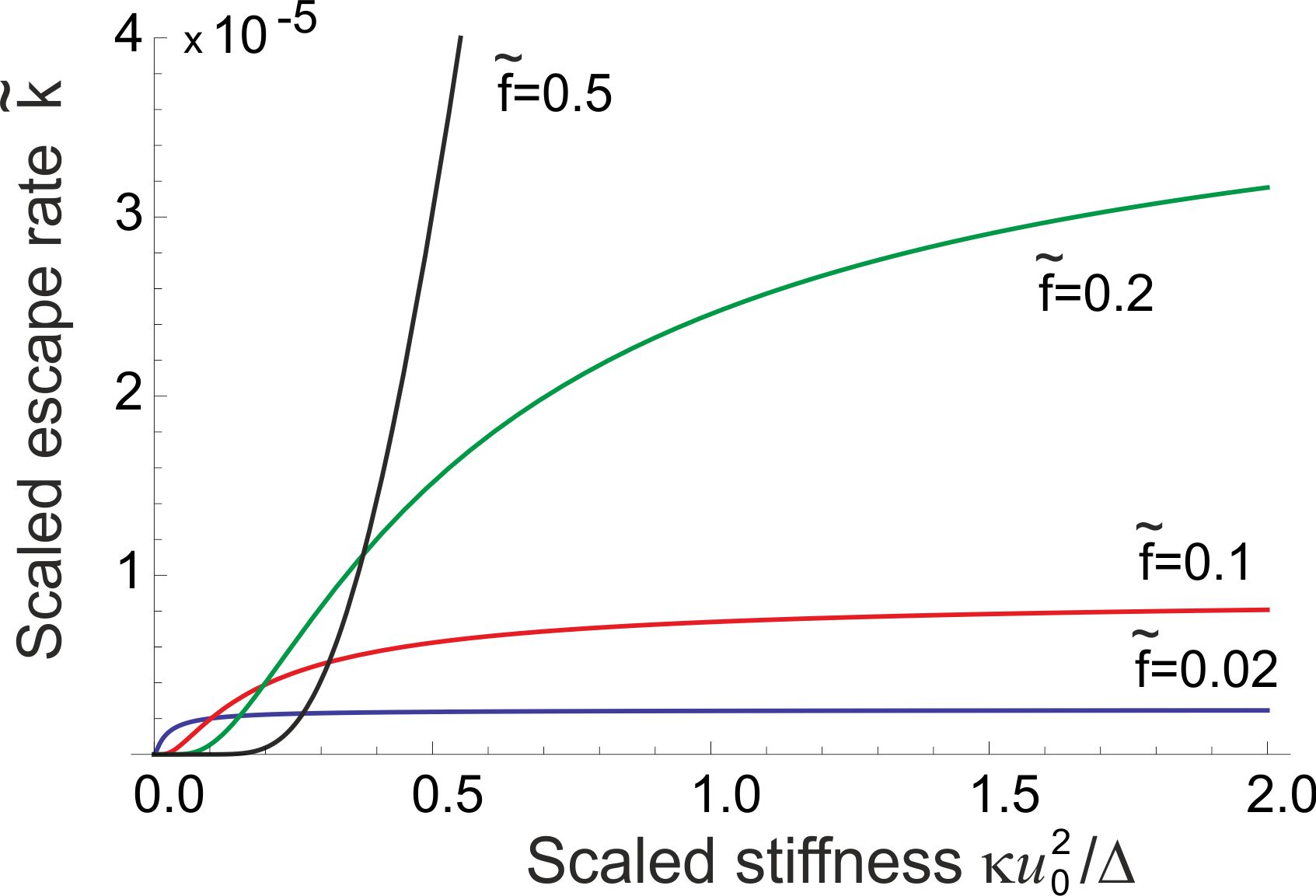}
\caption{The rate of latent complex rupture (and active TGF-$\beta$ escape), scaled in non-dimensional units as $\widetilde{k} = k\cdot (3\tilde{D}/2u_0^2) \sqrt{3 \Delta / 2\pi k_BT}$, plotted as function of substrate stiffness $\kappa$ (also scaled in non-dimensional units, see text). The curves are for three different values of scaled force: $\widetilde{f}=Fu_0/\Delta$, labeled on the plot. As the pulling force increases, the region of steepest response of the mechanosensing complex to the substrate stiffness shifts towards higher stiffness, and also becomes more diffuse -- enabling a graded response, as opposed to a very sharp response at small forces on very soft substrates.}
\label{fig:rateK}
\end{figure}

We begin by studying how the rate varies with different substrate stiffness, which was indeed the reason that stimulated this research in the first instance. Plotting the rate as a function of $\kappa$, fig.\,4, we observe that for any pulling force -- on stiffer substrates the latent complex is more likely to rupture, confirming the phenomenological interpretation discussed in the literature reviewed above. The functional form for the rate also appears to flatten considerably when $\kappa$ is greater than a certain threshold value, given by $\kappa^* \approx F^2/k_BT$, that is, both the crossover stiffness and the plateau value of the release rate are strong functions of the pulling force the cell exerts at this instance.
When $\kappa < \kappa^*$, small changes in stiffness will cause large changes in the TGF-$\beta$ release rate providing a positive feedback loop to further increase the $\alpha$-SMA, and ultimately, the strength of the pulling force. Above this value of stiffness, the sensors are telling the cell: `the substrate is as stiff as I can measure at the level of force I can offer'. When the force is very small -- most substrates will appear as ``stiff'': the cell is working in the saturated part of the curve $k(\kappa)$ with a weak feedback loop. On the other hand, if the pulling force is high to begin with, then very soft substrates offer no resistance and no TGF-$\beta$ is released to remodel the cell.

So one could trace gradual evolution of a cell after its first deposition onto a substrate from solution. Initially the cell applies a small pulling force $F$ and, unless the substrate is extremely soft (i.e. even at this initial force, $\kappa < \kappa^*$, when no response would follow), the rate of latent complex breaking takes an almost constant value $\widetilde{k} \approx 0.4$ (see the lowest curve in fig.\,4).  As the released TGF-$\beta$ gets absorbed on the cell surface, more $\alpha$-SMA is produced, stress fibers start to form and the pulling force $F$ increases. This further increases the rate of TGF-$\beta$ release, which is what experiments report.

The same process is presented from a different perspective in fig.\,5, where the rate of latent complex breaking is plotted for a given substrate (fixed $\kappa$) against the increasing force $F$. Apart from the case of very weak substrate, the rate grows as the cell remodels itself to increase the pulling force, until a maximum is reached. A further increase of force results in the negative feedback, which settles the cell evolution at homeostasis (a position labeled by * in figs.\,5 and 6). The stiffer the substrate, the higher the level of free TGF-$\beta$ and, accordingly, the amount of $\alpha$-SMA stress fibers one would find in this adjusted cell. Figure 6 focuses on the region of higher stiffness, where we find that the homeostatic point eventually disappears, replaced by the monotonic increase of the rupture rate with the applied force, all the way till the critical point $F_C = 3\Delta/2u_0$.

\begin{figure} 
\centering
\includegraphics[width=0.45\textwidth]{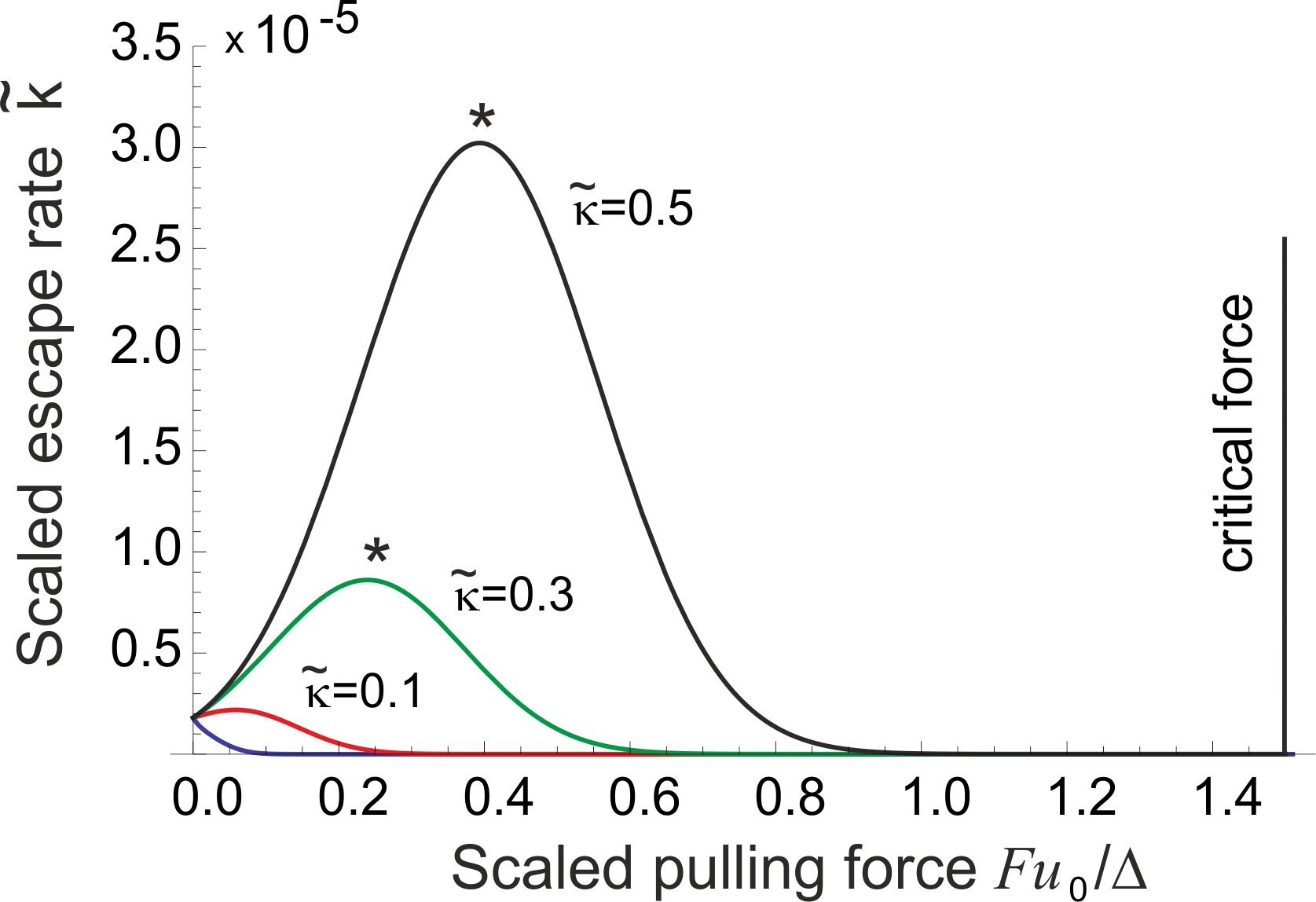}
\caption{The scaled rate of latent complex rupture (and active TGF-$\beta$ escape), $\widetilde{k}$, plotted as function of the applied pulling force (also scaled in non-dimensional units, see text). The curves are for different values of substrate stiffness: $\widetilde{\kappa}=\kappa u_0^2/\Delta$, labeled on the plot (the lowest curve barely visible at low forces is for $\widetilde{\kappa}=0.02$). On a stiffer substrate, there is a continuous increase in the escape rate of TGF-$\beta$ as the cell applies more force -- until a homeostatic point $F^*$ is reached (the maximum of $\kappa(F)$). Beyond that point the further increase of forces causes the negative feedback in the free TGF-$\beta$ release. On soft substrates, only a correspondingly weak pulling force illicits any response from the mechanosensing complex, with only the negative feedback. }
\label{fig:rateF}
\end{figure}

\begin{figure} 
\centering
\includegraphics[width=0.45\textwidth]{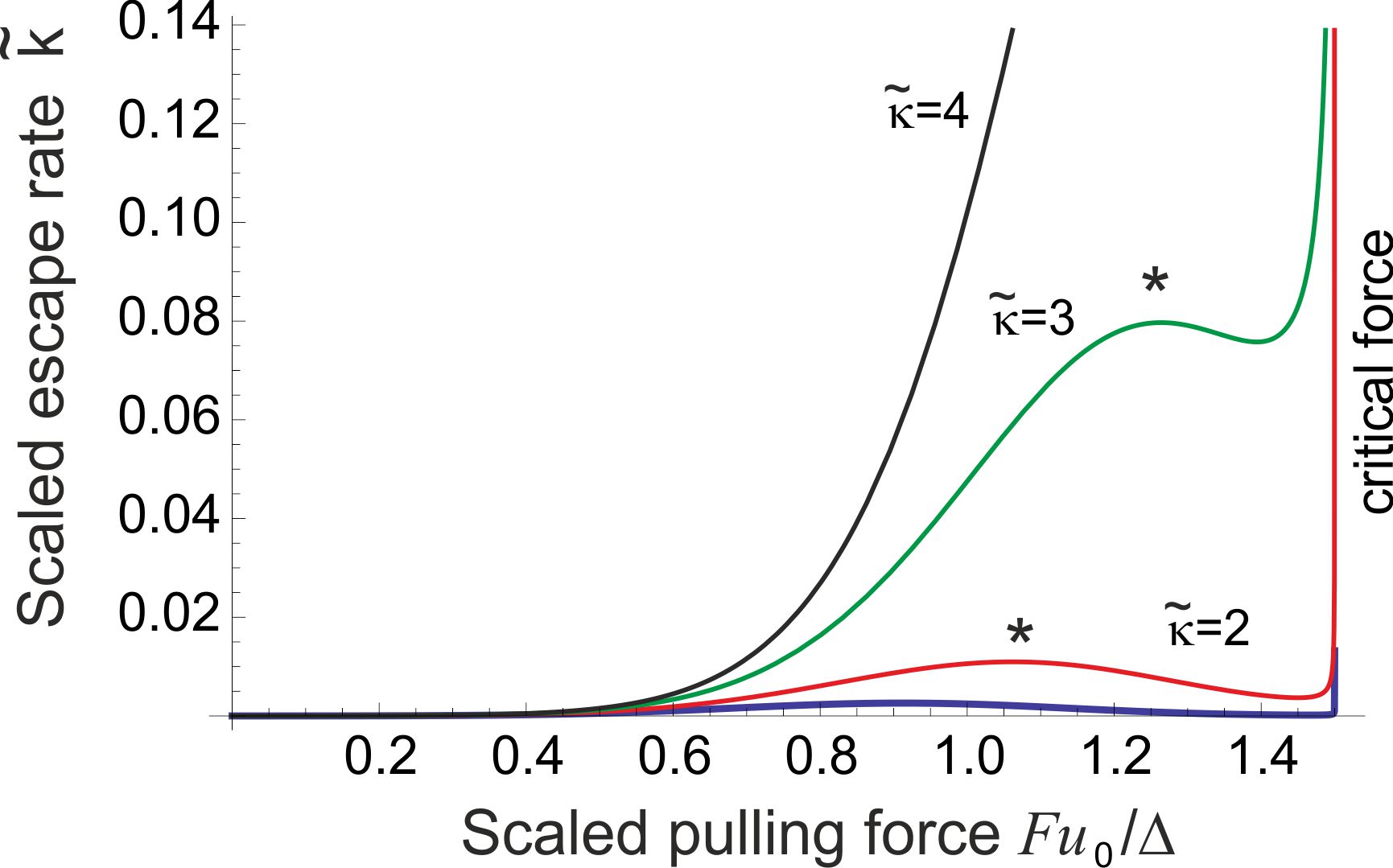}
\caption{The scaled rate of escape as function of the scaled applied pulling force (the same plot as fig.\,5) for a more stiff substrates.  The values of $\widetilde{\kappa}$ and the position of homeostatis ($F^*$) are labeled on the plot (the lowest barely visible curve is for $\widetilde{\kappa}=1$). On a very stiff substrate, $\widetilde{\kappa} > 3$, there is a continuous increase in the escape rate of TGF-$\beta$ as the cell applies more force with no homeostatic point $F^*$ (the maximum of $k(F)$) present: there is no region of negative feedback in the free TGF-$\beta$ release. The critical force limit $F=F_C$ is at $\widetilde{f} = 3/2$, at which point the latent complex is ruptured directly.  }
\label{fig:rateF}
\end{figure}

\section{The weak force approximation}

It is important to estimate the actual values of forces and energies involved. Although such an estimate is inevitably very inaccurate, depending on many individual factors of particular cells and conditions, the order of magnitude predictions have to be meaningful. Taking the characteristic energy scale of the latent complex $\Delta = 10$\,kcal/mol, the binding energy of a few hydrogen bonds; also $10-16$\,kcal/mol quoted by (29), and the distance needed to distort it to achieve rupture $u_0 = 0.3$\,nm (the size of a typical aminoacid monomer), we find the critical force $F_C = 350$\,pN. This is a high force unlikely to be generated by a single stress fiber of a cell. For comparison, the force to unfold integrin is quoted as 165\,pN (22). Buscemi et al. also quote $40$\,pN as the force specifically required to unfold the latent complex of TGF-$\beta$1 (22), although this is not purely a dynamic issue and at different times the unfolding would occur at different forces. Several other reports investigate the force required to disrupt the fibronectin-integrin-cytoskeleton linkage, finding the value of 1-2 pN (30,31). Various reports (14-16) have measured the actual pulling force $F$ exerted by different cells and in various circumstances, and on very different substrates -- nevertheless all converging on values between a few to a few tens of pN, see a summary review (32); note that a single myosin motor exerts about 3\,pN of force (33).

Note that an increasing body of experimental work finds (by using several ingenious techniques) that cells actually exert much greater overall forces on the substrate features. For instance, individual vascular smooth muscle cells, endothelial cells and fibroblasts exerted as much as 75\,nN of force (34), and forces at the edge of a sheet of epithelial cells were around 12\,nN, and around 5\,nN underneath the sheet (35). Lecuit et al.~(31) specifically report that focal adhesions a few micrometers long can bear forces up to 100\,nN.  This suggests to us that these high values of force refer to the large assembled (collective) constructions, whereas our interest here is a single molecular adhesion complex, with pN-level forces. We can expect a weak force at first, evolving to perhaps a much greater force on a stiff substrate when larger focal adhesions assemble, but this would serve as a reference point. For $F=10$\,pN, the scaled non-dimensional parameter used in plotting is $\widetilde{f} = F u_0/\Delta \sim 0.04$.

If a half-space occupied by an elastic medium (e.g. gel substrate) of the Young modulus $E$ is pulled along the free surface by a point force, another classical problem of Lord Kelvin (36), the response coefficient (spring constant) that we called the stiffness $\kappa$ is given by $\kappa = (4/3)\pi E \, \xi$ where $\xi$ is a short-distance cutoff: essentially the size of a microscopic object in the gel that the force is applied to. This sort of relation has been quoted by many authors, e.g. see the review (37) on surface forces in soft polymers. The only relevant length scale in our case (see fig.\,1) is the size of the attachment protein segment, which happens to be the same as a characteristic mesh size of a gel network, with an order of magnitude estimate of $\xi \sim 10$\,nm. For a very weak gel with $E = 10$\,kPa this gives the effective spring constant  $\kappa = 4.2 \cdot 10^{-4}$\,N/m and the non-dimensional parameter $\widetilde{\kappa} =\kappa u_0^2/\Delta \sim 5 \cdot 10^{-4}$. On stiff glass with $E=10$GPa, producing $\widetilde{\kappa} \sim 500$, so the whole spectrum of values is explored in a typical experiment.

From the estimates made above, however crude and generic they might be, it is clear that we can safely explore the limit of weak pulling force, $F/F_C \ll 1$. This allows writing down much simplified analytical expressions and expose the relevant factors contributing to the rate of breaking of latent TGF-$\beta$ complex on soft thermally-fluctuating substrates. This rate, expressed as $k=J/N_A$ with the parts separately given in Eqs.6 and 7 (see Supporting Information for more detail), takes the approximate form:
\begin{equation} \label{eq:rateSmall}
k \approx \frac{3\tilde{D}}{2u_0^2}  \sqrt{\frac{3 \Delta}{2\pi k_BT}} \left[1+\widetilde{f}\left(\beta{\Delta} -\frac{1}{\widetilde{\kappa}} \right)\right] e^{-\beta \Delta} e^{-\beta \Delta \frac{\widetilde{f}^2}{2\widetilde{\kappa}}},
\end{equation}
where $\beta \Delta $ is a shorthand for the ratio $(\Delta / k_BT)$ and the non-dimensional combinations $\widetilde{f}$ and $\widetilde{\kappa}$ measuring the force and the stiffness, respectively, have been defined above.  At vanishing pulling force, $F\rightarrow 0$, this expression reduces to a classical thermal activation rate, $k(0) \propto e^{-\beta \Delta}$, and the ratio $\beta \Delta $ has to be large for the latent complex not to fall apart spontaneously. Indeed the estimates above suggest that  $\beta \Delta \sim 16$.  The factor in square brackets in Eq.8 is the `remainder' of the Bell formula in the weak force approximation: if we ignore the soft substrate and take $\kappa \rightarrow \infty$, then this factor reduces to $ \exp[F u_0/k_BT]$. However, we now see that on a soft substrate the pulling force has a very different effect. By carefully analyzing the behavior of the function $k(F,\kappa)$ in different regimes of its parameters, we finally arrive at an interpolating expression that is very accurate in the weak force limit, and is almost exact in the limit of low stiffness -- yet remains practical and easy to manipulate. Here we present it in proper dimensional form, instead of using scaled $\widetilde{f}$ and $\widetilde{\kappa}$:
\begin{equation} \label{eq:rateInt}
k \approx \frac{3\tilde{D}}{2u_0^2}  \sqrt{\frac{3 \Delta}{2\pi k_BT}}\,  \exp \left[- \frac{\Delta -Fu_0}{ k_BT} - \frac{F}{\kappa u_0} - \frac{F^2}{2\kappa k_BT} \right].
\end{equation}

Equation 9 is the main analytical result of this paper. We remind that the effective stiffness parameter $\kappa$ is linearly proportional to the Young modulus of the substrate gel.  From here we can quickly estimate the substrate stiffness, below which the mechanosensing feedback is always negative (the low curve in fig.\,5) and thus we must assume the cell `rejects' the substrate of excessive softness and detaches from it: this negative slope of $k(F)$ at $F \rightarrow 0$ begins at $\widetilde{\kappa}_\textrm{min} \leq k_BT/\Delta$, or in proper dimensional units: $\kappa_\textrm{min} \leq k_BT/ u_0^2$. This critical stiffness value is purely determined by the level of thermal noise in the substrate. The other important threshold is the stiffness of a substrate, above which there is no longer a homeostatic point, i.e. the increase of the rate of escape $k(F)$ becomes monotonic (see fig.\,6). The point of $\widetilde{\kappa} \approx 3$ corresponds to $\kappa = 0.02$\,N/m and that occurs at the Young modulus of $E \approx 0.6$\,MPa. This is a modulus of a typical tough rubber, so one can see that the cell mechanosensing response on a solid substrate (including the bone, where the modulus measures in GPa range) should be completely different from the response on soft gels (including muscle or even brain tissue).

Equations~8 and 9, although a strong approximation, allow us to analytically describe homeostasis on soft substrates. This is a very important point, as here the cell remodeling in response to the TGF-$\beta$ mechanosensing signalling should stop: any further increase of free TGF-$\beta$, the resulting production of $\alpha$-SMA and the increase of the pulling force will cause negative feedback and return the cell to this homeostasis position. The pulling force at this point is easily obtained by finding the maximum of $k(F)$ in Eq.~9, although we must be aware that this is only a qualitative estimate -- the full solution for a maximum of $k(F)=J/N_A$ is cumbersome and best studied numerically; in proper dimensional units we have:
\begin{equation} \label{eq:Fstar}
F^* = \frac{k_BT}{u_0}    \left( \frac{\kappa u_0^2}{k_BT} - 1\right),
\end{equation}
where we could employ $\kappa/\kappa_\textrm{min}$ as a shorthand for the ratio $\kappa u_0^2/k_BT$, which essentially compares the elastic and the thermal energy scales in  the substrate (instead of the earlier used $\widetilde{\kappa} = \kappa u_0^2/\Delta$ that compares the elasticity of substrate and of the latent complex). Again we see that thermal fluctuations, represented by the energy factor $k_BT$, are at the core of mechanosensing homeostasis.
Therefore, with our chosen parameters (for TGF-$\beta$ mechanism in (myo)fibroplasts and smooth muscle cells) the homeostasis exists between $\kappa_\textrm{min}$ and $\kappa_\textrm{max} \approx 28 \kappa_\textrm{min}$, at which point $F^* \geq F_C$. That is, for stiff substrates with a Young modulus $E$ above 30\,MPa the cell will no longer be able to achieve homeostasis and will continue to produce stress fibers to increase its pulling force. We might speculate that this is a point of coronary disease onset when the SMC's in arterial wall detect a stiff cholesterol plaque underneath.  The much lower range of homeostatic substrate stiffness for the neurons, and the higher homeostatic range for osteoblatsts arise because of a different value of ``lock'' stiffness $\Delta$ (using a different variant of TGF-$\beta$ or even a different latent complex altogether) and a different value of lock deformation threshold $u_0$.

 One must be conscious that the theoretical model developed here simplifies many complicated factors and overlooks many details, only focusing on the essential mechanism. Our aim was not to achieve quantitative accuracy, which requires a much more careful input from the biological model, in particular, into the form of the ``lock potential'' $U(x)$ and the way the protein binds to the gel substrate (which determines $\kappa$) -- but to offer an internally self-consistent and mathematically non-controversial model of mechanosensitivity to replace the earlier more vague ideas.

\section{Conclusions}\label{Conclusions}
This work was inspired by a critical question about one of the most fundamental interactions within living organisms. When trying to explain the mechanical interactions of cells with their environments an apparent paradox is encountered in the impossibility of measuring stiffness by a probe without a reference point. An extensive literature on this subject carefully sidestepped this issue. Our work shows that even a simple model for a biomolecular mechanosensor can overcome this paradox by using the Brownian motion in the substrate as the `reference reservoir': the answer lies in stochastic forces prominent in biomolecular systems. The sensor has evolved in such a way that it is extremely sensitive to the thermally-fluctuating behavior of its environment and is able to extract useful information from it. Different substrate stiffness changes not only the equilibrium configuration of the mechanosensor, but also various features in its characteristic potential landscape.

The proposed model for a biomolecular mechanosensor is a likely candidate for several similar sensors. We have described the system as a sensor under constant tension force in series with a viscoelastic substrate, where signalling occurs via a change in the sensor's configuration. We derived an analytical expression for the expected rate of rupture of the latent complex as a function of applied force and substrate elasticity. The final expression for $k(F,\kappa)$ provides convincing behavior in all regimes. It gives the experimentally observed dependence on the substrate stiffness $\kappa$: a higher rate of signalling with increasing $\kappa$, with a highly nonlinear dependence with clear biological advantages. It also shows that the $\kappa$-dependence may be broken down in two distinct regimes, of an extreme sensitivity to stiffness and of saturation and weak $\kappa$ dependence,  with the crossover point dependent on the magnitude of pulling force  $F$: another aspect that matches experimental observation and physical intuition. The expression for the rate also shows a qualitatively satisfactory dependence on $F$. The rate of signalling increases with $F$ as the potential barrier to be overcome is lowered -- finds a point of homeostasis where the cell attempts to return to from both directions -- and then steeply rises to infinity as the external force increases much further towards  the critical point, for which the barrier is destroyed and both the model and the latent complex break down. Finally the expression obtained reduces to the Bell solution in the high barrier limit, a `sanity-check' result which matches many simple experiments.

The model developed here builds on and improves related work in the literature, yet it is the first one to use analytical stochastic methods of solution applied to the problem of mechanosensitivity. It extends the model proposed by (13) in a different context to incorporate a substrate with separate mechanical and stochastic characteristics, and shows the correct behavior in the vicinity of the critical force.

In recent years, among the large literature on this subject of mechanical forces felt and exerted by cells, there has been a suggestion that perhaps it is not the constant force, but a fixed deformation of the substrate that cells feel (38,39). We find this concept difficult to reconcile with what we understand about the action of myosin motors on cytoskeleton and the mechanical force balance in quasi-equilibrium the adhesion point must be in. There are other experiments that specifically state that the contractile force generated by the fibroblasts was independent of the stiffness of the resistance (40), and we wish to side with this view.

One can see the analogy of the proposed stochastic mechanism for the change in breaking rate of the TGF-$\beta$ latent complex on soft substrates with the principle of the enzymatic action (41). The accelerated rate of reaction is achieved by an enzyme localizing the reacting particle near the site (via positional and rotational constraints). In our case, for any given force $F$ from the cell, the localization in the potential well (A) is more pronounced -- and so the rate of escape over the barrier increased (even though the height of the barrier itself, $\Delta V$ is not actually changed). In contrast, on a soft substrate, a shallower potential well results in a lower confinement and correspondingly -- a lower rate of the latent complex breaking. One may say that the thermal motion is more readily transferred into a random motion of the attachment point in a soft substrate, and so makes a lesser contribution to the dissociation process.

Further work on this system could include a generalization of the rate expression to a more complex and realistic form for the latent complex potential. It would also be interesting to explore how numerical values obtained by substituting typical cellular pulling forces and distances compare with experiment: we used a very generic set of values while there are no doubt many variations in specific cases.

\subsection*{Acknowledgments}
This work was funded by the Sims Scholarship, the Cambridge Trusts, and the University of Sydney. We are grateful for many useful discussions with  K. Franze, B. Hinz, U. Schwarz  and S. Sinha.

\end{document}